\documentclass[a4paper,11pt]{article}
\pdfoutput=1

\usepackage{jheppub}
\usepackage{graphicx}
\usepackage{ifthen}

\newcommand{\Dmq}{\Delta m^2}
\newcommand{\eVq}{\ensuremath{\text{eV}^2}}
\newcommand{\Nuc}[2][]{{\ensuremath{\ifthenelse{\equal{#1}{}}{}{\mbox{}^{#1}}\text{#2}}}}
\newcommand{\Gauss}{\mathop{\mathrm{Gauss}}}
\newcommand{\Lnot}{{\;\slash\!\!\!\!L}}

\makeatletter
\DeclareRobustCommand\recite[1]{\begingroup\@fileswfalse\cite{#1}\endgroup}
\makeatother

\allowdisplaybreaks

\title{Updated Determination of the Solar Neutrino Fluxes from
  Solar Neutrino Data}

\author[a]{Johannes Bergstr\"om,}
\affiliation[a]{Departament d'Estructura i Constituents de la
  Mat\`eria and Institut de Ciencies del Cosmos, Universitat de
  Barcelona, Diagonal 647, E-08028 Barcelona, Spain}
\emailAdd{bergstrom@ecm.ub.edu}

\author[a,b,c]{M.~C.~Gonzalez-Garcia,}
\affiliation[b]{Instituci\'o Catalana de Recerca i Estudis
  Avan\c{c}ats (ICREA)}
\affiliation[c]{C.N.~Yang Institute for Theoretical Physics, State
  University of New York at Stony Brook, Stony Brook, NY 11794-3840,
  USA}
\emailAdd{maria.gonzalez-garcia@stonybrook.edu}

\author[d]{Michele Maltoni,}
\affiliation[d]{Instituto de F\'{\i}sica Te\'orica UAM/CSIC, Calle de
  Nicol\'as Cabrera 13--15, Universidad Aut\'onoma de Madrid,
  Cantoblanco, E-28049 Madrid, Spain}
\emailAdd{michele.maltoni@csic.es}

\author[e]{Carlos Pe\~na-Garay,}
\affiliation[e]{Instituto de F\'{\i}sica Corpuscular (IFIC), CSIC \&
  Universitat de Valencia Calle Catedr\'atico Jos\'e Beltr\'an, 2,
  E-46090 Paterna, Valencia, Spain}
\emailAdd{penya@ific.uv.es}

\author[f]{Aldo M. Serenelli,}
\affiliation[f]{nstitut de Ciencies de l'Espai (ICE-CSIC/IEEC), Campus
  UAB, Carrer de Can Magrans s/n, 08193 Cerdanyola del Valls, Spain}
\emailAdd{aldos@ice.csic.es}

\author[c]{Ningqiang Song,}
\emailAdd{ningqiang.song@stonybrook.edu}

\abstract{We present an update of the determination of the solar
  neutrino fluxes from a global analysis of the solar and terrestrial
  neutrino data in the framework of three-neutrino mixing. Using a
  Bayesian analysis we reconstruct the posterior probability
  distribution function for the eight normalization parameters of the
  solar neutrino fluxes plus the relevant masses and mixing, with and
  without imposing the luminosity constraint. We then use these
  results to compare the description provided by different Standard
  Solar Models. Our results show that, at present, both models with
  low and high metallicity can describe the data with equivalent
  statistical agreement. We also argue that even with the present
  experimental precision the solar neutrino data have the potential to
  improve the accuracy of the solar model predictions.}

\preprint{IFT-UAM/CSIC-15-135, YITP-SB-16-1}

\keywords{solar neutrinos}

\begin{document}

\maketitle

\section{Introduction}

The Sun generates power through nuclear fusion, the basic energy
source being the conversion of four protons into an alpha particle,
two positrons and two neutrinos. As early as 1939~\cite{Bethe:1939bt},
Bethe identified two different mechanisms by which such overall
process could take place, now known as the pp-chain and the
CNO-cycle~\cite{Bahcall:1989ks}. In the pp-chain, fusion reactions
among elements lighter than $A = 8$ produce a characteristic set of
neutrino fluxes, whose spectral energy shapes are known but whose
normalization must be calculated with a detailed solar model.  In the
CNO-cycle the abundance of \Nuc[12]{C} plus \Nuc[13]{N} acts as a
catalyst, while the \Nuc[13]{N} and \Nuc[15]{O} beta decays provide
the primary source of neutrinos.

In order to precisely determine the rates of the different reactions
in the two chains and to obtain the final neutrino fluxes and their
energy spectrum, a detailed modeling of the Sun is needed. Standard
Solar Models (SSMs)~\cite{Bahcall:1987jc, TurckChieze:1988tj,
  Bahcall:1992hn, Bahcall:1995bt, Bahcall:2000nu, Bahcall:2004pz,
  PenaGaray:2008qe, Serenelli:2011py} derive the properties of the
present Sun by following its evolution after entering the main
sequence. The models use as inputs a set of observational parameters
(the present surface abundances of heavy elements and surface
luminosity of the Sun, as well as its age, radius and mass) and rely
on some basic assumptions: spherical symmetry, hydrostatic
equilibrium, initial homogeneous composition, evolution at constant
mass.  Over the past five decades the solar models were steadily
refined with the inclusion of more precise observational and
experimental information about the input parameters (such as nuclear
reaction rates and the surface abundances of different elements), with
more accurate calculations of constituent quantities (such as
radiative opacity and equation of state), the inclusion of new
physical effects (such as element diffusion), and the development of
faster computers and more precise stellar evolution codes.

The produced neutrinos, given their weak interactions, can exit the
Sun practically unaffected, and therefore enable us to see into the
solar interior and verify directly our understanding of the
Sun~\cite{Bahcall:1964gx}. This was the goal of the original solar
neutrino experiments, which was somewhat diverted by the appearance of
the so-called ``solar neutrino problem''~\cite{Bahcall:1968hc,
  Bahcall:1976zz}. Such problem has now been fully solved through the
modification of the Standard Model with inclusion of neutrino masses
and mixing, which allow for flavor transition of the neutrino from
production to detection~\cite{Pontecorvo:1967fh, Gribov:1968kq,
  Wolfenstein:1977ue, Mikheev:1986gs} and for non-trivial effects (the
so called LMA-MSW flavor transitions) when crossing dense regions of
matter.  The upcoming of the real-time experiments Super-Kamiokande
and SNO and the independent determination of the flavor oscillation
probabilities using reactor antineutrinos at KamLAND has allowed for
the precise determination of the neutrino parameters (masses and
mixing) responsible for these flavor transitions.

In parallel to the increased precision in our understanding of
neutrino propagation, a new puzzle has emerged in the consistency of
SSMs~\cite{Bahcall:2004yr}.  SSMs built in the 1990's were very
successful in predicting other observations. In particular, quantities
measured by helioseismology such as the radial distributions of sound
speed and density~\cite{Bahcall:1992hn, Bahcall:1995bt,
  Bahcall:2000nu, Bahcall:2004pz} showed good agreement with the
predictions of the SSM calculations and provided accurate information
on the solar interior.  A key element to this agreement is the input
value of the abundances of heavy elements on the surface of the
Sun~\cite{Grevesse:1998bj}. However, since 2004 new determinations of
these abundances have become available, pointing towards substantially
lower values~\cite{Asplund:2004eu, Asplund:2009fu}. The SSMs based on
such lower metallicities fail at explaining the helioseismic
observations~\cite{Bahcall:2004yr}.

So far there has not been a successful solution of this puzzle as
changes in the Sun modeling do not seem able to account for this
discrepancy~\cite{Castro:2007, Guzik:2010, Serenelli:2011py}.  Thus
the situation is that, at present, there is no fully consistent SSM.
This led to the construction of two different sets of SSMs, one based
on the older solar abundances~\cite{Grevesse:1998bj} implying high
metallicity, and one assuming lower metallicity as inferred from more
recent determinations of the solar abundances~\cite{Asplund:2004eu,
  Asplund:2009fu}. In Ref.~\cite{Serenelli:2009yc, Serenelli:2011py}
the solar fluxes corresponding to such two models were detailed, based
on updated versions of the solar model calculations presented in
Ref.~\cite{Bahcall:2004pz}.

In Ref.~\cite{GonzalezGarcia:2009ya} we performed a solar model
independent analysis of the solar and terrestrial neutrino data in the
framework of three-neutrino masses and mixing, aiming at
simultaneously determine the flavor parameters and all the solar
neutrino fluxes with a minimum set of theoretical priors. Since then
more data have been accumulated by the solar neutrino experiments, and
new non-solar neutrino experiments have provided a more accurate
determination of the neutrino oscillation parameters.  Thus in this
work we present an update of our former analysis. In
Sec.~\ref{sec:ana} we briefly summarize our methodology, data included
and physical assumptions. In Sec.~\ref{sec:res} we give the new
reconstructed posterior probability distribution function for the
eight normalization parameters of the solar neutrino fluxes, with and
without the constraint imposed by the observed solar luminosity.  In
Sec.~\ref{sec:compamod} we use the results of this analysis to
statistically test to what degree the present solar neutrino data can
discriminate between the two SSMs, and we estimate whether the present
data are precise enough to provide useful information to the
construction of the SSM.  Finally in Sec.~\ref{sec:sum} we summarize
our conclusions.

\section{Analysis framework}
\label{sec:ana}

In the analysis of solar neutrino experiments we include the total
rates from the radiochemical experiments
Chlorine~\cite{Cleveland:1998nv}, Gallex/GNO~\cite{Kaether:2010ag} and
SAGE~\cite{Abdurashitov:2009tn}. For real-time experiments we include
the results on electron scattering (ES) from the four phases in
Super-Kamiokande: the 44 data points of the phase I (SK1)
energy-zenith spectrum~\cite{Hosaka:2005um}, the 33 data points of the
full energy and day/night spectrum in phase II
(SK2)~\cite{Cravens:2008aa}, the 42 energy and day/night data points
in phase III (SK3)~\cite{Abe:2010hy}, and the 24 data points of the
energy spectrum and day-night asymmetry of the 1669-day of phase IV
(SK4)~\cite{sksol:nu2014}.  The results of the three phases of SNO are
included in terms of the parametrization given in their combined
analysis~\cite{Aharmim:2011vm} which amount to 7 data points.  We also
include the main set of the 740.7 days of Borexino Phase-1
data~\cite{Bellini:2011rx, Bellini:2013lnn} as well as their
high-energy spectrum from 246 live days~\cite{Bellini:2008mr} and the
408 days of Borexino Phase-2 data~\cite{Bellini:2014uqa}.  Details of
our Borexino Phase-2 data analysis which is totally novel in this
article are presented in Appendix~\ref{sec:app-borex}.
In the framework of three neutrino masses and mixing the expected
values for these solar neutrino observables depend on the parameters
$\Dmq_{21}$, $\theta_{12}$, and $\theta_{13}$ as well as on the
normalizations of the eight solar fluxes.

Besides solar experiments, we also include the observed energy
spectrum in KamLAND data sets DS-1 and DS-2~\cite{Gando:2010aa} with a
total exposure of $3.49\times 10^{32}$ target-proton-year (2135 days),
which in the framework of three neutrino mixing also yield information
on the parameters $\Dmq_{21}$, $\theta_{12}$, and $\theta_{13}$.

In addition, we include the information on $\theta_{13}$ obtained
after marginalizing over $\Dmq_{3\ell}$, $\theta_{23}$ and
$\delta_\textsc{cp}$ the results of all the other oscillation
experiments considered in the NuFIT-2.0 analysis presented in
Refs.~\cite{Gonzalez-Garcia:2014bfa, Bergstrom:2015rba, nufit}. In
particular this accounts for Super-Kamiokande atmospheric neutrino
data from phases SK1--4~\cite{skatm:nu2014} (with addition of the 1775
days of phase SK4 over their published results on phases
SK1--3~\cite{Wendell:2010md}); the energy distribution of long
baseline neutrinos from MINOS in both $\nu_\mu$ and $\bar\nu_\mu$
disappearance with $10.71 \times 10^{20}$ and $3.36 \times
10^{20}$~pot, respectively, as well as from T2K in $\nu_\mu$
disappearance~\cite{Abe:2014ugx} with $6.57\times 10^{20}$ pot; LBL
appearance results from MINOS~\cite{Adamson:2013ue} with exposure
$10.6 \times 10^{20}$ ($\nu_e$) and $3.3 \times 10^{20}$ ($\bar\nu_e$)
pot, and from T2K with $6.57\times 10^{20}$ pot
($\nu_e$)~\cite{Abe:2013hdq}; reactor data from the finalized
experiments CHOOZ~\cite{Apollonio:1999ae} and Palo
Verde~\cite{Piepke:2002ju} together with the spectrum from Double
Chooz with 227.9 days live time~\cite{Abe:2012tg}, and the 621-day
spectrum from Daya Bay~\cite{db:nu2014}, as well as the near and far
rates observed at RENO with 800 days of
data-taking~\cite{reno:nu2014}.

\begin{table}
  \catcode`?=\active\def?{\hphantom{0}}
  \begin{tabular}{r@{\hspace{20mm}}c@{\hspace{20mm}}c@{\hspace{20mm}}c}
    Flux & $\Phi_i^\text{ref}$ [$\text{cm}^{-2}\, \text{s}^{-1}$]
    & $\alpha_i$ [MeV] & $\beta_i$
    \\
    \hline
    \Nuc{pp}    & $5.98\times 10^{10}$ & $13.0987$ & $9.186\times 10^{-1}$ \\
    \Nuc[7]{Be} & $5.00\times 10^{9?}$ & $12.6008$ & $7.388\times 10^{-2}$ \\
    \Nuc{pep}   & $1.44\times 10^{8?}$ & $11.9193$ & $2.013\times 10^{-3}$ \\
    \Nuc[13]{N} & $2.96\times 10^{8?}$ & $?3.4577$ & $1.200\times 10^{-3}$ \\
    \Nuc[15]{O} & $2.23\times 10^{8?}$ & $21.570?$ & $5.641\times 10^{-3}$ \\
    \Nuc[17]{F} & $5.52\times 10^{6?}$ & $?2.3630$ & $1.530\times 10^{-5}$ \\
    \Nuc[8]{B}  & $5.58\times 10^{6?}$ & $?6.6305$ & $4.339\times 10^{-5}$ \\
    \Nuc{hep}  & $8.04\times 10^{3?}$ & $?3.7370$ & $3.523\times 10^{-8}$
  \end{tabular}
  \caption{The reference neutrino flux $\Phi_i^\text{ref}$ used for
    normalization, the energy $\alpha_i$ provided to the star by
    nuclear fusion reactions associated with the $i^\text{th}$
    neutrino flux (taken from Ref.~\recite{Bahcall:2001pf}), and the
    fractional contribution $\beta_i$ of the $i^\text{th}$ nuclear
    reaction to the total solar luminosity.}
  \label{tab:lumcoef}
\end{table}

In what follows, for convenience, we will use as normalization
parameters for the solar fluxes the reduced quantities:
\begin{equation}
  \label{eq:redflux}
  f_i = \frac{\Phi_i}{\Phi_i^\text{ref}}
\end{equation}
with $i = \Nuc{pp}$, \Nuc[7]{Be}, \Nuc{pep}, \Nuc[13]{N},
\Nuc[15]{O}, \Nuc[17]{F}, \Nuc[8]{B}, and \Nuc{hep}. The numerical
values of $\Phi_i^\text{ref}$ are set to the predictions of the GS98
solar model as given in Ref.~\cite{Serenelli:2011py} and are listed in
Table~\ref{tab:lumcoef}.
With this, the theoretical predictions for the relevant observables
(after marginalizing over $\Dmq_{23}$, $\theta_{23}$ and
$\delta_\textsc{cp}$) depend on eleven parameters: the three relevant
oscillation parameters $\Dmq_{21}$, $\theta_{12}$, $\theta_{13}$ and
the eight reduced solar fluxes $f_i$.
With the data from the different data samples (D) and the theoretical
predictions for them in terms of these parameters $\vec\omega =
(\Dmq_{21}, \theta_{12}, \theta_{13}, f_{\Nuc{pp}}, \dots,
f_{\Nuc{hep}})$ we build the corresponding likelihood function
\begin{equation}
  \mathcal{L}(\mathrm{D} | \vec\omega)
  = \frac{1}{N} \exp\left[ - \frac{1}{2} \chi^2(\mathrm{D} |
    \vec\omega) \right]
\end{equation}
where $N$ is a normalization factor.
In Bayesian statistics our knowledge of $\vec\omega$ is summarized by
the posterior probability distribution function (pdf)
\begin{equation}
  \label{eq:ppdf}
  p(\vec\omega|\mathrm{D},\mathcal{P}) =
  \dfrac{\mathcal{L}(\mathrm{D} | \vec\omega)\,
    \pi(\vec\omega | \mathcal{P})}{\mathcal{Z}_\mathcal{P}}
\end{equation}
where in the denominator we have introduced the so-called
\emph{evidence} ${\mathcal {Z}}_{\mathcal{P}}$
\begin{equation}
  \label{eq:evdef}
  \mathcal {Z}_\mathcal{P} \equiv \Pr(D|\mathcal{P})
  = \int \mathcal{L}(\mathrm{D} | \vec\omega')\,
  \pi(\vec\omega' | \mathcal{P})\, d\vec\omega'
\end{equation}
which gives the likelihood for the hypothesis (or model) $\mathcal{P}$
to describe the data. Here $\pi(\vec\omega | \mathcal{P})$ is the
prior probability density for the parameters in the hypothesis
$\mathcal{P}$.

In our model-independent analysis we assume a uniform prior
probability complemented by a set of constraints to ensure consistency
in the pp-chain and CNO-cycle, as well as some relations from nuclear
physics. Specifically, we impose the following restrictions:
\begin{itemize}
\item The fluxes must be positive:
  \begin{equation}
    \label{eq:fpos}
    \Phi_i \geq 0 \quad\Rightarrow\quad f_i \geq 0 \,.
  \end{equation}

\item The number of nuclear reactions terminating the pp-chain should
  not exceed the number of nuclear reactions which initiate
  it~\cite{Bahcall:1995rs, Bahcall:2001pf}:
  \begin{multline}
    \Phi_{\Nuc[7]{Be}} + \Phi_{\Nuc[8]{B}}
    \leq \Phi_{\Nuc{pp}} + \Phi_{\Nuc{pep}}
    \\
    \Rightarrow\quad
    8.49 \times 10^{-2} f_{\Nuc[7]{Be}}
    + 9.95 \times 10^{-5} f_{\Nuc[8]{B}}
    \leq f_{\Nuc{pp}} + 2.36 \times 10^{-3} f_{\Nuc{pep}} \,.
  \end{multline}

\item The $\Nuc[14]{N}(p,\gamma) \Nuc[15]{O}$ reaction must be the
  slowest process in the main branch of the
  CNO-cycle~\cite{Bahcall:1995rs}:
  \begin{equation}
    \label{eq:CNOineq1}
    \Phi_{\Nuc[15]{O}} \leq \Phi_{\Nuc[13]{N}}
    \quad\Rightarrow\quad
    f_{\Nuc[15]{O}} \leq 1.34 f_{\Nuc[13]{N}}
  \end{equation}
  and the CNO-II branch must be subdominant:
  \begin{equation}
    \label{eq:CNOineq2}
    \Phi_{\Nuc[17]{F}} \leq \Phi_{\Nuc[15]{O}}
    \quad\Rightarrow\quad
    f_{\Nuc[17]{F}}\leq 37 f_{\Nuc[15]{O}} \,.
  \end{equation}

\item The ratio of the \Nuc{pep} neutrino flux to the \Nuc{pp}
  neutrino flux is fixed to high accuracy because they have the same
  nuclear matrix element. We have constrained this ratio to match the
  average of the GS98 and AGSS09 values, with $1\sigma$ Gaussian
  uncertainty given by the difference between the values in the two
  models\footnote{We have verified that assuming a flat distribution
    over the $1\sigma$ uncertainty interval does not produce
    significant differences in the results of our analysis.}
  \begin{equation}
    \label{eq:pep-pp}
    \frac{f_{\Nuc{pep}}}{f_{\Nuc{pp}}} = 1.006 \pm 0.013 \,.
  \end{equation}
\end{itemize}
In this work we use \textsc{MultiNest}~\cite{Feroz:2007kg,
  Feroz:2008xx, Feroz:2013hea}, a Bayesian inference tool which, given
the prior and the likelihood, calculates the evidence with an
uncertainty estimate, and generates posterior samples from
distributions that may contain multiple modes and pronounced (curving)
degeneracies in high dimensions.

The number of independent fluxes is reduced when imposing the
so-called ``luminosity constraint'', \textit{i.e.}, the requirement
that the sum of the thermal energy generation rates associated with
each of the solar neutrino fluxes coincides with the solar
luminosity~\cite{Spiro:1990vi}:
\begin{equation}
  \label{eq:lumsum}
  \frac{L_\odot}{4\pi \, (\text{A.U.})^2}
  = \sum_{i=1}^8 \alpha_i \Phi_i \,.
\end{equation}
Here the constant $\alpha_i$ is the energy provided to the star by the
nuclear fusion reactions associated with the $i^\text{th}$ neutrino
flux; its numerical value is independent of details of the solar model
to an accuracy of one part in $10^4$ or better~\cite{Bahcall:2001pf}.
A detailed derivation of this equation and the numerical values of the
coefficients $\alpha_i$, which we reproduce for convenience in
Table~\ref{tab:lumcoef}, is presented in Ref.~\cite{Bahcall:2001pf}.
In terms of the reduced fluxes Eq.~\eqref{eq:lumsum} can be written
as:
\begin{equation}
  1 = \sum_{i=1}^8 \beta_i f_i
  \quad\text{with}\quad
  \beta_i \equiv
  \frac{\alpha_i \Phi_i^\text{ref}}{L_\odot \big/ [4\pi \, (\text{A.U.})^2]}
\end{equation}
where $\beta_i$ is the fractional contribution to the total solar
luminosity of the nuclear reactions responsible for the production of
the $\Phi_i^\text{ref}$ neutrino flux, and $L_\odot \big/ [4\pi \,
  (\text{A.U.})^2] = 8.5272 \times 10^{11} \, \text{MeV} \,
\text{cm}^{-2} \, \text{s}^{-1}$~\cite{Bahcall:2001pf}.
The analysis performed incorporating the priors in
Eqs.~(\ref{eq:fpos}--\ref{eq:lumsum}) will be named ``analysis with
luminosity constraint'', $\mathcal{P} = L_\odot$, and for this case
the prior probability distribution is:
\begin{equation}
  \pi(\vec\omega' | L_\odot) =
  \begin{cases}
    \dfrac{1}{N} \exp\left[
      -\dfrac{\left(f_{\Nuc{pep}} \big/ f_{\Nuc{pp}} - 1.006 \right)^2}{2\sigma^2}
      \right]
    & \text{if Eqs.~(\ref{eq:fpos}--\ref{eq:CNOineq2})
      and~\eqref{eq:lumsum} are verified,}
    \\
    0 & \text{otherwise,}
  \end{cases}
\end{equation}
where $N$ is a normalization factor and $\sigma=0.010$.  When only
Eqs.~(\ref{eq:fpos}--\ref{eq:pep-pp}) are imposed we will speak of
``analysis without luminosity constraint'', $\mathcal{P} =
\Lnot_\odot$, so:
\begin{equation}
  \pi(\vec\omega' | {\Lnot_\odot}) =
  \begin{cases}
    \dfrac{1}{N} \exp\left[
      -\dfrac{\left(f_{\Nuc{pep}} \big/ f_{\Nuc{pp}} - 1.006 \right)^2}{2\sigma^2}
      \right]
    & \text{if Eqs.~(\ref{eq:fpos}--\ref{eq:CNOineq2}) are verified,}
    \\
    0 & \text{otherwise.}
  \end{cases}
\end{equation}
Let us notice that the conditions in
Eqs.~(\ref{eq:fpos}--\ref{eq:CNOineq2}) and Eq.~\eqref{eq:lumsum} are
constraints on some linear combinations of the solar fluxes and they
are model independent, \textit{i.e.}, they do not impose any prior
bias favoring either of the SSMs. Furthermore we have chosen to
center the condition~\eqref{eq:pep-pp} at the average of the GS98 and
AGSS09 values, with $1\sigma$ Gaussian uncertainty given by the
difference between the values in the two models, to avoid the
introduction of a bias towards one of the models.  In the next
sections we will comment on how our results are affected when this
prior is centered about the GS98 or the AGSS09 prediction.


\section{Determination of solar neutrino fluxes}
\label{sec:res}

\begin{figure}\centering
  \includegraphics[width=0.95\textwidth]{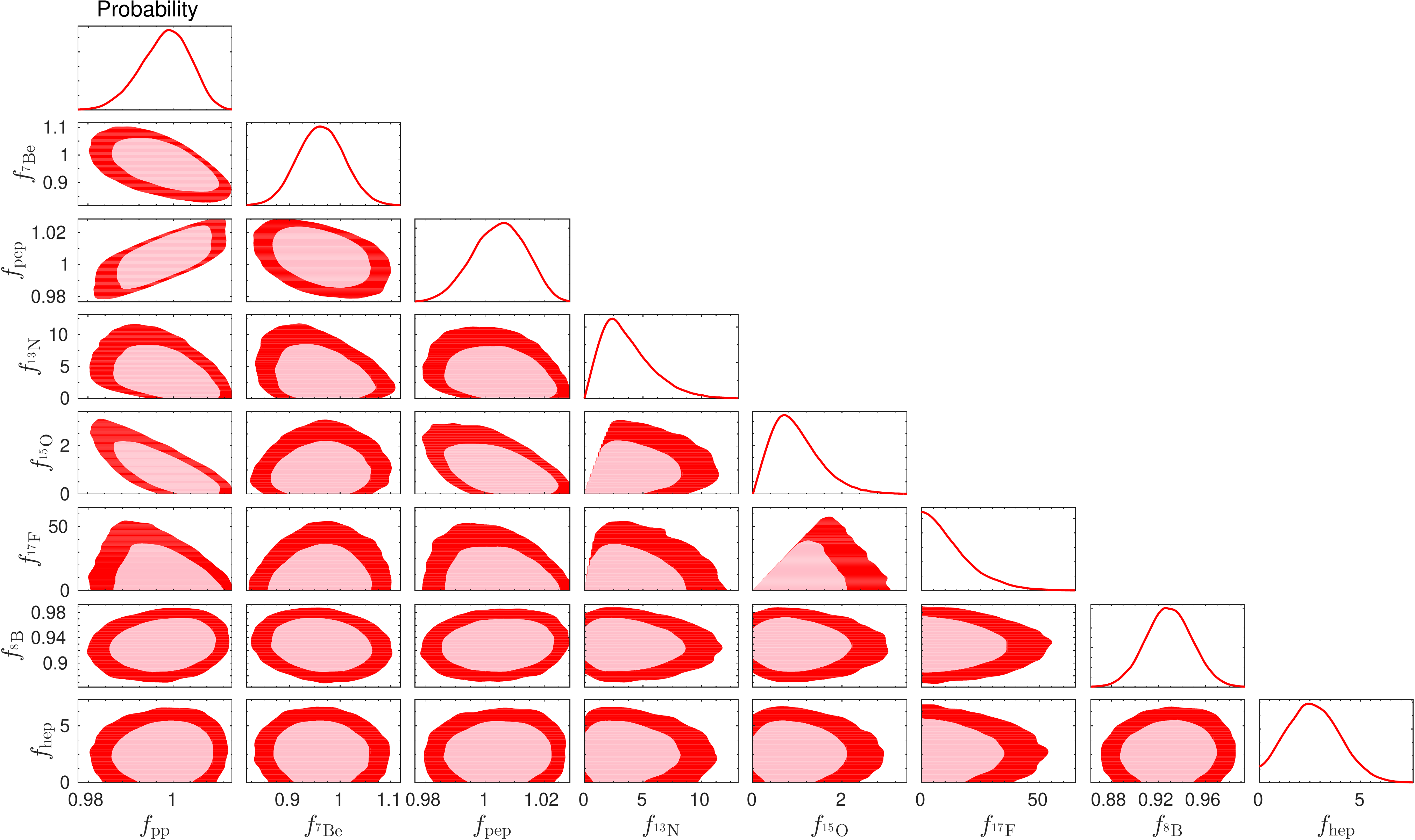}
  \caption{Constraints from our global analysis on the solar neutrino
    fluxes. The curves in the right-most panels show the marginalized
    one-dimensional probability distributions. The rest of the panels
    show the 90\% and 99\% CL two-dimensional credibility regions (see
    text for details).}
  \label{fig:LCtri}
\end{figure}

Our results for the analysis with luminosity constraint are displayed
in Fig.~\ref{fig:LCtri}, where we show the marginalized
one-dimensional probability distributions $p(f_i | \mathrm{D},
L_\odot) $ for the eight solar neutrino fluxes as well as the 90\% and
99\% CL two-dimensional allowed regions. The corresponding ranges at
$1\sigma$ (and at the 99\% CL in square brackets) on the oscillation
parameters are:
\begin{equation}
  \label{eq:bestosc}
  \begin{aligned}
    \Dmq_{21}
    &= 7.5 \pm 0.2 \, [^{+0.4}_{-0.5}]  \times 10^{-5}~\eVq \,,
    \\
    \sin^2\theta_{12}
    &= 0.30 \pm 0.01 \, [^{+0.04}_{-0.03}] \,,
    \\
    \sin^2\theta_{13}
    &= 0.022 \pm 0.001 \, [^{+0.002}_{-0.003}] \,.
  \end{aligned}
\end{equation}
while for the solar neutrino fluxes we get:
\begin{equation}
  \label{eq:bestlc}
  \begin{aligned}
    f_{\Nuc{pp}}
    & = 0.999 ^{+0.006}_{-0.005} \, [^{+0.012}_{-0.016}] \,, \qquad
    & \Phi_{\Nuc{pp}}
    & = 5.971 ^{+0.037}_{-0.033} [^{+0.073}_{-0.097}]
    \times 10^{10}~\text{cm}^{-2}~\text{s}^{-1} \,,
    \\
    f_{\Nuc[7]{Be}}
    & = 0.96 ^{+0.05}_{-0.04} \, [^{+0.12}_{-0.11}] \,, \qquad
    & \Phi_{\Nuc[7]{Be}}
    & = 4.80^{+0.24}_{-0.22} \, [^{+0.60}_{-0.57}]
    \times 10^{9}~\text{cm}^{-2}~\text{s}^{-1} \,,
    \\
    f_{\Nuc{pep}}
    & = 1.005 \pm0.009 \, [^{+0.019}_{-0.024}] \,, \qquad
    & \Phi_{\Nuc{pep}}
    & = 1.448 \pm0.013 \, [^{+0.028}_{-0.034}]
    \times 10^{8}~\text{cm}^{-2}~\text{s}^{-1} \,,
    \\
    f_{\Nuc[13]{N}}
    & = 1.7^{+2.9}_{-1.0} \, [^{+8.4}_{-1.6}] \,, \qquad
    & \Phi_{\Nuc[13]{N}}
    & \leq 13.7 \, [30.2]
    \times 10^{8}~\text{cm}^{-2}~\text{s}^{-1} \,,
    \\
    f_{\Nuc[15]{O}}
    & = 0.6 ^{+0.6}_{-0.4} \, [^{+2.0}_{-0.6}] \,, \qquad
    & \Phi_{\Nuc[15]{O}}
    & \leq 2.8\, [5.8]
    \times 10^{8}~\text{cm}^{-2}~\text{s}^{-1} \,,
    \\
    f_{\Nuc[17]{F}}
    & \leq 15 \, [46] \,, \qquad
    & \Phi_{\Nuc[17]{F}}
    & \leq 8.5 \, [25]
    \times 10^{7}~\text{cm}^{-2}~\text{s}^{-1} \,,
    \\
    f_{\Nuc[8]{B}}
    & = 0.92 \pm0.02 \, [\pm 0.05] \,, \qquad
    & \Phi_{\Nuc[8]{B}}
    & = 5.16^{+0.13}_{-0.09} \, [^{+0.30}_{-0.26}]
    \times 10^{6}~\text{cm}^{-2}~\text{s}^{-1} \,,
    \\
    f_{\Nuc{hep}}
    & = 2.4^{+1.5}_{-1.2} \, [\leq 5.9 ] \,, \qquad
    & \Phi_{\Nuc{hep}}
    & = 1.9 ^{+1.2}_{-0.9} \, [\leq 4.7]
    \times 10^{4}~\text{cm}^{-2}~\text{s}^{-1} \,.
  \end{aligned}
\end{equation}

We notice that with the exception of $\Nuc[17]{F}$ all other fluxes
have a vanishing (or close to) probability for their corresponding
$f=0$.  However, it is important to stress that for what concerns
$f_{\Nuc[13]{N}}$ and $f_{\Nuc[15]{O}}$ this is mostly consequence of
the inequalities in Eqs.~\eqref{eq:CNOineq1} and~\eqref{eq:CNOineq2},
which effectively result into priors behaving as $\pi(f_i) \propto
f_i$ for small $f_i$. For this reason the corresponding $1\sigma$
credible intervals for these fluxes, constructed as iso-posterior
intervals and shown in the left column of Eq.~\eqref{eq:bestlc}, do
not extend to $f_i = 0$ even though setting $f_{\Nuc[13]{N}} =
f_{\Nuc[15]{O}} = f_{\Nuc[17]{F}} = 0$ gives a reasonable fit to the
data. With this in mind, in the right column in Eq.~\eqref{eq:bestlc}
we have chosen to quote only the $1\sigma$ and 99\%CL upper boundaries
for the corresponding solar neutrino fluxes, rather than the complete
allowed range.

As mentioned above we have checked the stability of the results under
changes in the assumption of the Gaussian prior in
Eq.~\eqref{eq:pep-pp}.  We find that if we center this prior at the
GS98 prediction ($f_{\Nuc{pep}} \big/ f_{\Nuc{pp}} = 1$) the best fit
value for $\Nuc{pep}$ neutrinos is changed to $f_{\Nuc{pep}} = 0.998$
($\Phi_{\Nuc{pep}} = 1.437\times 10^{8}\, \text{cm}^{-2}\,
\text{s}^{-1}$), while if we center it at the AGSS09 prediction
($f_{\Nuc{pep}} \big/ f_{\Nuc{pp}} = 1.013$) we get $f_{\Nuc{pep}} =
1.012$ ($\Phi_{\Nuc{pep}} = 1.457\times 10^{8}\, \text{cm}^{-2}\,
\text{s}^{-1}$). All other fluxes are unaffected.

\begin{figure}\centering
  \includegraphics[width=0.95\textwidth]{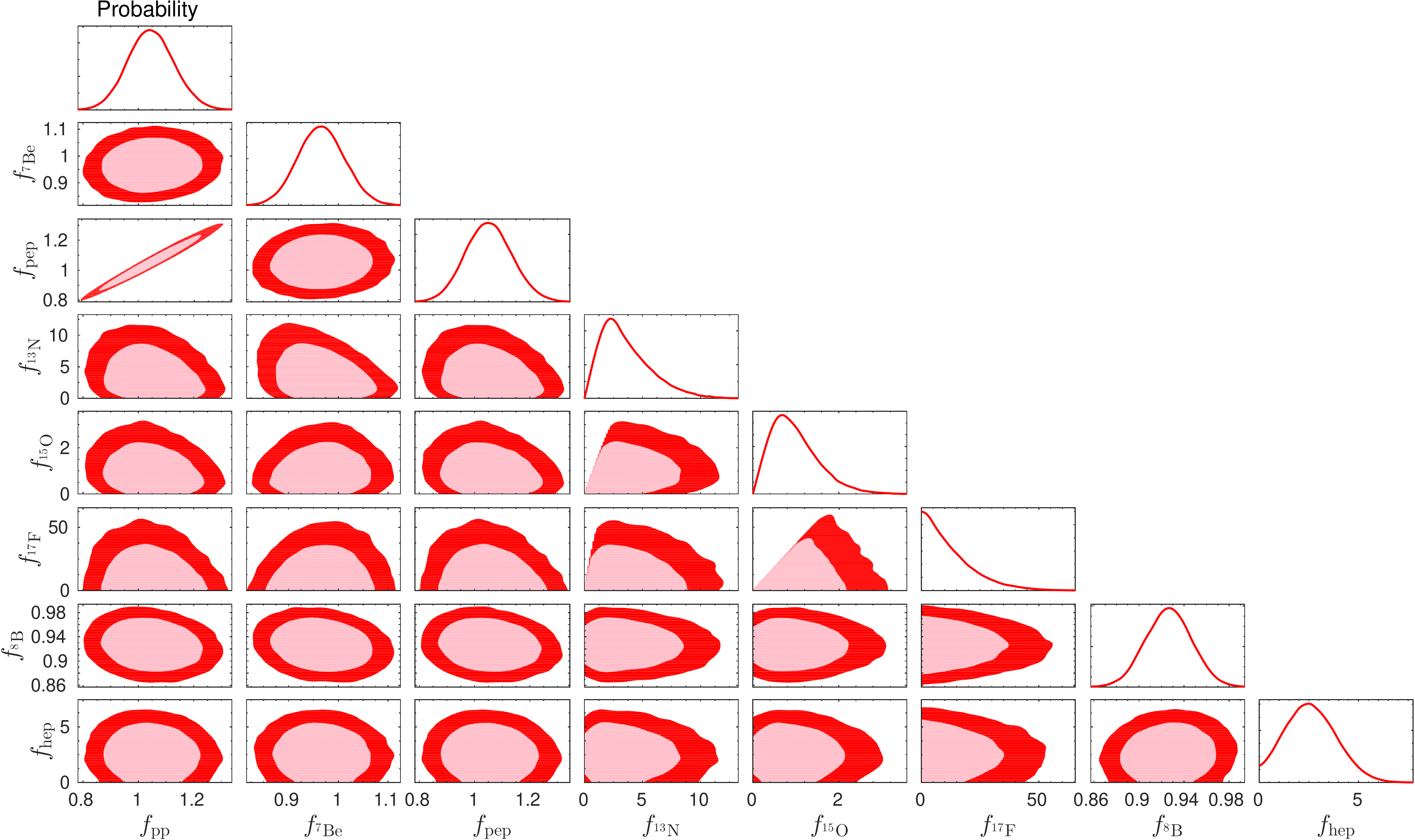}
  \caption{Same as Fig.~\ref{fig:LCtri} but without the luminosity
    constraint, Eq.~\eqref{eq:lumsum}.}
  \label{fig:NoLCtri}
\end{figure}

As seen in Fig.~\ref{fig:LCtri} the most important correlation appears
between the \Nuc{pp} and \Nuc{pep} fluxes, as expected from the
relation~\eqref{eq:pep-pp}. The correlation between the \Nuc{pp} (and
\Nuc{pep}) and \Nuc[7]{Be} flux is directly dictated by the
luminosity constraint (see comparison with Fig.~\ref{fig:NoLCtri}).
All these results imply the following share of the energy production
between the pp-chain and the CNO-cycle
\begin{equation}
  \label{eq:ppcnolum1}
  \frac{L_\text{pp-chain}}{L_\odot} =
  0.991 ^{+0.005}_{-0.004} \, [^{+0.008}_{-0.013}]
  \quad\Longleftrightarrow\quad
  \frac{L_\textsc{cno}}{L_\odot} =
  0.009 ^{+0.004}_{-0.005} \, [^{+0.013}_{-0.008}] \,,
\end{equation}
in perfect agreement with the SSMs which predict $L_\textsc{cno} /
L_\odot \leq 1$\% at the $3\sigma$ level. Note that the same comment
as on the $f_{\Nuc[13]{N}}$ and $ f_{\Nuc[15]{O}}$ fluxes applies to
the total CNO luminosity, so we can understand the result in
Eq.~\eqref{eq:ppcnolum1} effectively as an upper bound on the
contribution of the CNO-cycle to the Sun Luminosity: ${L_\textsc{cno}}
/ {L_\odot} \leq 2.2$\% at 99\% CL.

In order to check the consistency of our results we have performed the
same analysis without imposing the luminosity constraint,
Eq.~\eqref{eq:lumsum}. The corresponding results for $p(f_i |
\mathrm{D}, \Lnot_\odot)$ and the two-dimensional allowed regions are
shown in Fig.~\ref{fig:NoLCtri}.
As expected, the \Nuc{pp} flux is the most affected by the release of
this constraint. This is so because the \Nuc{pp} reaction gives the
largest contribution to the solar energy production, as can be seen in
Table~\ref{tab:lumcoef}. Hence, using the luminosity constraint only
as an upper bound would imply that the \Nuc{pp} flux cannot exceed its
SSM prediction by more than 9\%, while completely removing this
constraint allows for a much larger \Nuc{pp} flux.  The \Nuc{pep} flux
is also severely affected due to its strong correlation with the
\Nuc{pp} flux, Eq.~\eqref{eq:pep-pp}.  On a smaller scale the CNO
fluxes are also affected, mainly as an indirect effect due to the
modified contribution of the \Nuc{pp} and \Nuc{pep} fluxes to the
Gallium and Chlorine experiments, which leads to a change in the
allowed CNO contribution to these experiments. Thus in this case we
get:
\begin{equation}
  \begin{aligned}
    \label{eq:bestnolc}
    f_{\Nuc{pp}}
    & = 1.04 \pm 0.08 \, [^{+0.22}_{-0.20}]  \,,
    \\
    f_{\Nuc[7]{Be}}
    &= 0.97^{+0.04}_{-0.05} \, [\pm0.12] \,,
    \\
    f_{\Nuc{pep}}
    & = 1.05 \pm0.08 \, [^{+0.23}_{-0.20}]  \,,
    \\
    f_{\Nuc[13]{N}}
    &= 1.7^{+2.8}_{-1.0} \, [^{+8.4}_{-1.6}] \,,
    \\
    f_{\Nuc[15]{O}}
    &= 0.6 ^{+0.7}_{-0.4} \, [\leq 2.6] \,,
    \\
    f_{\Nuc[17]{F}}
    & \leq 15 \, [47] \,.
  \end{aligned}
\end{equation}
The determination of the \Nuc[8]{B} and \Nuc{hep} fluxes (as well as
the oscillation parameters) is basically unaffected by the luminosity
constraint.

Interestingly, the idea that the Sun shines because of nuclear fusion
reactions can be tested accurately by comparing the observed photon
luminosity of the Sun with the luminosity inferred from measurements
of solar neutrino fluxes. We find that the energy production in the
pp-chain and the CNO-cycle without imposing the luminosity constraint
are given by:
\begin{equation}
  \label{eq:ppcnolum2}
  \frac{L_\text{pp-chain}}{L_\odot}
  = 1.03^{+0.08}_{-0.07} \, [^{+0.21}_{-0.18}]
  \qquad\text{and}\qquad
  \frac{L_\textsc{cno}}{L_\odot}
  = 0.008^{+0.005}_{-0.004} \, [^{+0.014}_{-0.007}] \,.
\end{equation}
Comparing Eqs.~\eqref{eq:ppcnolum1} and~\eqref{eq:ppcnolum2} we see
that the luminosity constraint has only a limited impact on the amount
of energy produced in the CNO-cycle. However, as discussed above, the
amount of energy in the pp-chain can now significantly exceed the
total quantity allowed by the luminosity constraint.  Altogether we
find that the present value for the ratio of the neutrino-inferred
solar luminosity, $L_\odot \text{(neutrino-inferred)}$, to the photon
luminosity $L_\odot$ is:
\begin{equation}
  \label{eq:lnutot}
  \frac{L_\odot \text{(neutrino-inferred)}}{L_\odot}
  = 1.04 [^{+0.07}_{-0.08}] \, [^{+0.20}_{-0.18}] \,.
\end{equation}
Thus we find that, at present, the neutrino-inferred luminosity
perfectly agrees with the measured one, and this agreement is known
with a $1\sigma$ uncertainty of 7\%, which is a factor two smaller
than the previous best determination~\cite{GonzalezGarcia:2009ya}.

\section{Comparison with the Standard Solar Models}
\label{sec:compamod}

\begin{figure}\centering
  \includegraphics[width=0.95\textwidth]{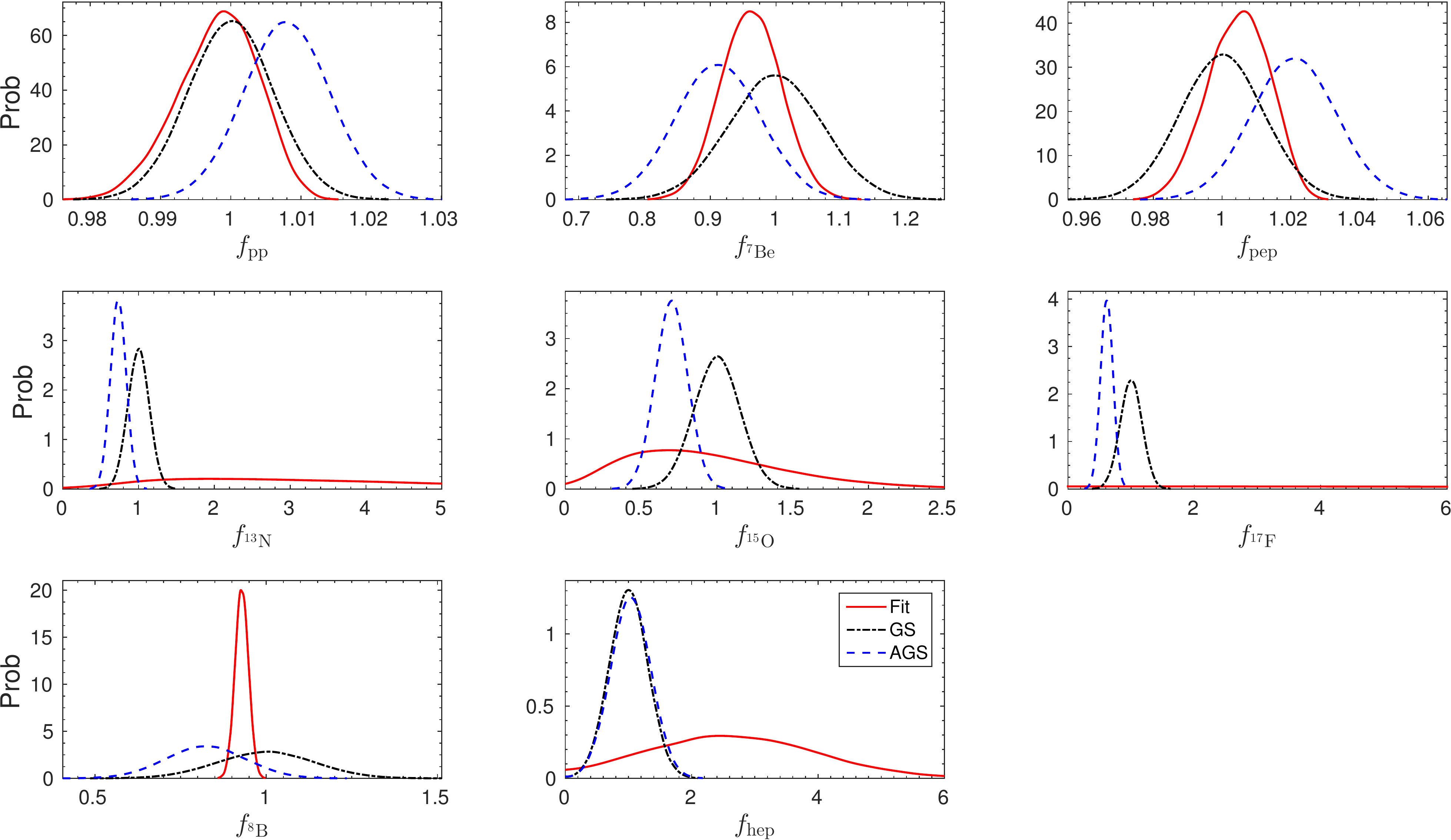}
  \caption{Marginalized one-dimensional probability distributions for
    the best determined solar fluxes in our analysis as compared to
    the predictions for the two SSMs in Ref.~\recite{Serenelli:2011py}.}
  \label{fig:SSMwflat}
\end{figure}

Next we compare the results of our determination of the solar fluxes
with the expectations from the solar models, SSM=GS (for GS98) and
SSM=AGS (for AGSS09).  In order to do so we use the predictions
$\langle f_i^\textsc{ssm} \rangle$ for the fluxes, the relative
uncertainties $\sigma_i^\textsc{ssm}$ and their correlations
$\rho_{ij}^\textsc{ssm}$ in both models as obtained from
Refs.~\cite{Serenelli:2011py, SSMweb}. The prior distribution
$\pi(\vec{f}|\text{SSM})$ with maximum entropy (\textit{i.e.}, minimum
information) satisfying these constraints is a multivariate normal
distribution, and this is what we assume in what follows. In
Fig.~\ref{fig:SSMwflat} we show the marginalized one-dimensional
probability distributions for the solar neutrino fluxes as determined
by our analysis including the luminosity constraint, together with the
corresponding prior distributions for the two SSMs.

\begin{table}\centering
  \begin{tabular}{|@{\hspace{10mm}}c@{\hspace{10mm}}|@{\hspace{10mm}}c@{\hspace{10mm}}|c|}
    \hline
    $\left|\log(\text{odds})\right|$ &  odds & Interpolation
    \\
    \hline
    $< 1.0$ & $\lesssim 3:1$ & Inclusive
    \\
    \hline
    1.0 & $\simeq 3:1$ & Weak evidence
    \\
    \hline
    2.5 & $\simeq 12:1$ & Moderate evidence
    \\
    \hline
    5.0 & $\simeq 150:1$ & Strong evidence
    \\
    \hline
  \end{tabular}
  \caption{Values of Jeffrey's scale used for the interpretation of
    model odds.}
  \label{tab:Jeffrey}
\end{table}

Comparison between the two models can be achieved by calculating the
posterior odds, given data $\mathrm{D}$, simply using Bayes' theorem
\begin{equation}
  \frac{\Pr(\text{GS}|\mathrm{D})}{\Pr(\text{AGS}|\mathrm{D})}
  = \frac{\Pr(\mathrm{D}|\text{GS})\,\pi(\text{GS})}{\Pr(\mathrm{D}|\text{AGS})\,\pi(\text{AGS})}
  = \frac{\mathcal{Z}_\textsc{gs}}{\mathcal{Z}_\textsc{ags}}
  \frac{\pi(\text{GS})}{\pi(\text{AGS})}
\end{equation}
where we compute the evidences $\mathcal{Z}_\textsc{ssm}$ as in
Eq.~\eqref{eq:evdef} with the prior distributions for the $f_i$ in
each model and taking $\pi(\text{GS})/\pi(\text{AGS})$, the prior probability
ratio for the two models, to be unity (this is, a priori both models
are taken to be equally probable).  The posterior odds can interpreted
using the Jeffreys scale in Table~\ref{tab:Jeffrey}.

Our calculation shows that $\log\mathcal{Z}_\textsc{gs} /
\mathcal{Z}_\textsc{ags} = 0.00 \pm 0.05$, meaning that the data has
\emph{absolutely} no preference to either model.  Quantitatively this
result is driven by the most precisely measured \Nuc[8]{B} flux,
which, as seen in Fig.~\ref{fig:SSMwflat}, lies right in the middle of
the predictions of GS98 and AGSS09.  In what respects the possible
discriminating power from the other precisely measured fluxes, in
particular \Nuc[7]{Be} and indirectly pp and pep, one must realize
that within the SSMs the fluxes originating from the pp-chain are
rather correlated among them; therefore, after the determination of
the \Nuc[8]{B} flux is imposed the posterior predictions of all the
other pp-chain fluxes are also \emph{pushed} towards the average of
the two models, essentially making them indistinguishable with respect
to measurements of these fluxes.
In order to estimate how the correlations predicted by the SSM affect
the comparison of the solar models, we define two new schemes
$\text{GS}'$ and $\text{AGS}'$ where such correlations have been
removed, \textit{i.e.}, $\rho_{ij}^\textsc{ssm}=\delta_{ij}$.  In this
case we find $\log\mathcal{Z}_{\textsc{gs}'} /
\mathcal{Z}_{\textsc{ags}'} = 0.2 \pm 0.1$, meaning that even without
the effect of the pp-chain correlations present data are unable to
break the degeneracy between models implied by the \Nuc[8]{B}
measurement.

\begin{figure}\centering
  \includegraphics[width=0.7\textwidth]{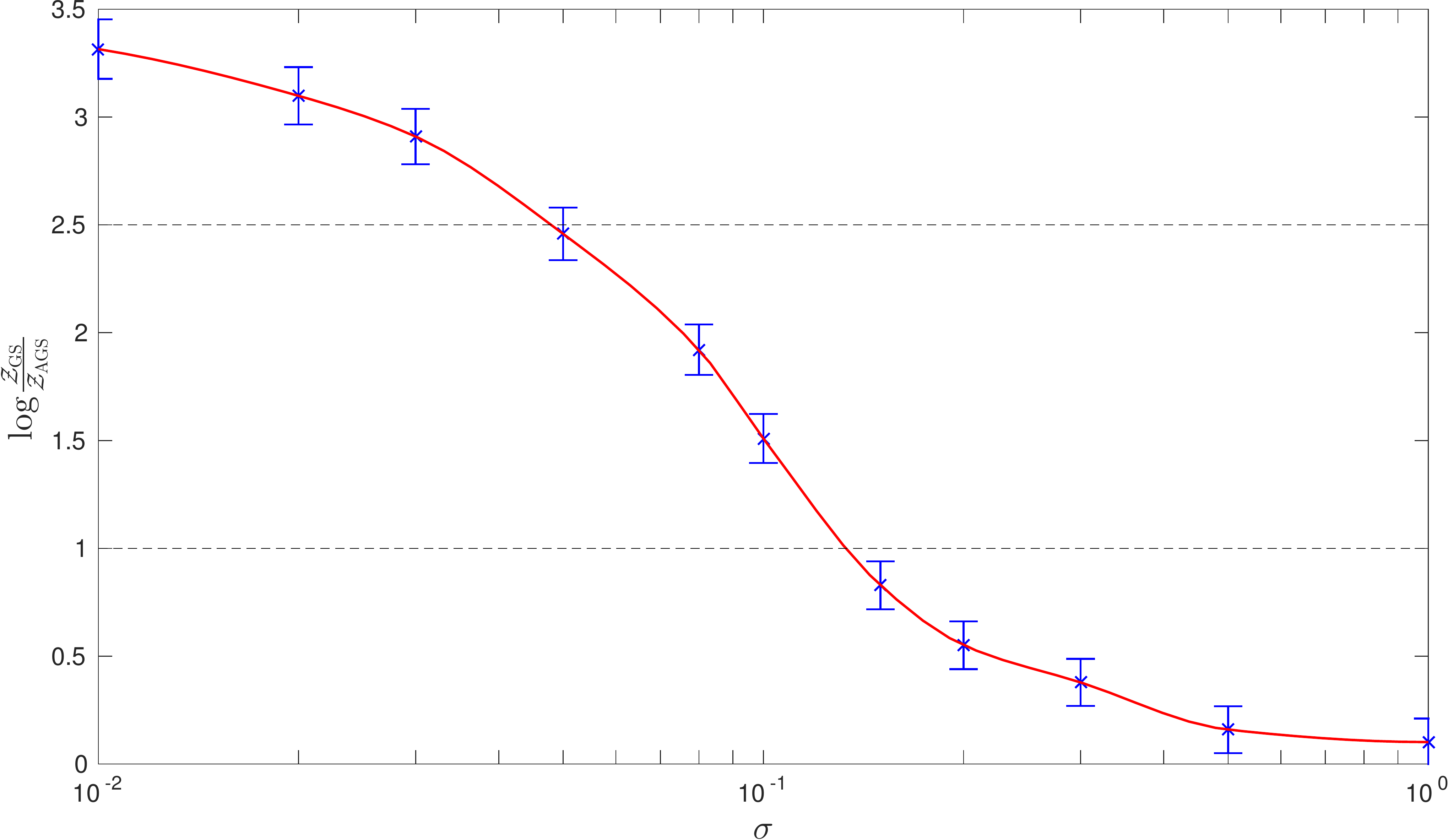}
  \caption{Bayes factor as a function of the assumed relative error on
    $\Phi_\textsc{cno}$. The bars give the numerical uncertainty of
    our calculations and the curve is a cubic interpolation. The
    dashed lines marks the limits for weak and moderate evidence of
    the Jeffreys scale, respectively.}
  \label{fig:testCNO}
\end{figure}

On the other hand, the CNO fluxes are rather uncorrelated with the
pp-chain fluxes, so even with the ``democratic'' \Nuc[8]{B} flux
result discussed above one could aim at discriminating between the
solar models by measuring the CNO fluxes (also taking into account
that their expectations strongly differ between the two models, as
seen Fig.~\ref{fig:SSMwflat}).  To quantify this possibility we repeat
our analysis including also an hypothetical future measurement of the
total CNO flux, $\Phi_\textsc{cno} = f_{\Nuc[13]{N}}
\Phi_{\Nuc[13]{N}}^\text{ref} + f_{\Nuc[15]{O}}
\Phi_{\Nuc[15]{O}}^\text{ref} + f_{\Nuc[17]{F}}
\Phi_{\Nuc[17]{F}}^\text{ref}$, characterized by a given uncertainty
$\sigma_\textsc{cno}$ and centered at the prior expectation of one of
the models (for example the GS98 model, $\hat{\Phi}_\textsc{cno} =
5.24\times 10^8\, \text{cm}^{-2}\, \text{s}^{-1}$). We plot in
Fig.~\ref{fig:testCNO} the result of this exercise where we show the
log of the Bayes factor as a function of the assumed relative error on
$\Phi_\textsc{cno}$. From this figure we read that within the present
model uncertainties a moderate evidence in favor of the model whose
CNO fluxes have been assumed (GS98 in this case) can be achieved by a
measurement of such fluxes with $\sigma_\textsc{cno} = 5\%$ accuracy.

\section{Generalizing/strengthening the solar models}
\label{sec:gen}

Finally we make a first attempt to address whether the present data is
precise enough to give relevant information which could be used as
input for the construction of a more robust SSM. In order to do so we
devise an analysis in which we naively generalize the SSM predictions
by two parameters which are meant to characterize the best SSM from
the point of view of the solar neutrino data.

First we notice that for most fluxes the theoretical correlations
between the flux predictions of the solar models are pointing ``in the
same direction'' as the difference between the mean of the predictions
of the models. So it seems reasonable to make the solar models
slightly more robust by letting the mean of the prediction vary
continuously as
\begin{equation}
  \label{eq:interp}
  \hat{f}(t) = t \hat{f}_\textsc{gs}
  + (1-t) \hat{f}_\textsc{ags},
\end{equation}
where $t$ now is an additional parameter. The AGS and GS solar models
are recovered for $t=0$ and $t=1$, respectively. Then, by calculating
the marginal likelihood of $t$, one can also evaluate the extent to
which either of the two solar models is preferred or not compared to
larger deviations (along the line of Eq.~\eqref{eq:interp}). In
addition, the Bayes factor calculated previously is simply the ratio
of the marginal likelihood at $t=0$ and $t=1$, which serves as an
additional check.

Second we consider how the inclusion of the neutrino data could affect
``on average'' the theoretical uncertainties of the model
predictions. In order to do so we introduce a second parameter
$\omega$ by which we rescale all $\sigma_i^\textsc{ssm}$.

\begin{figure}\centering
  \includegraphics[width=1.\textwidth]{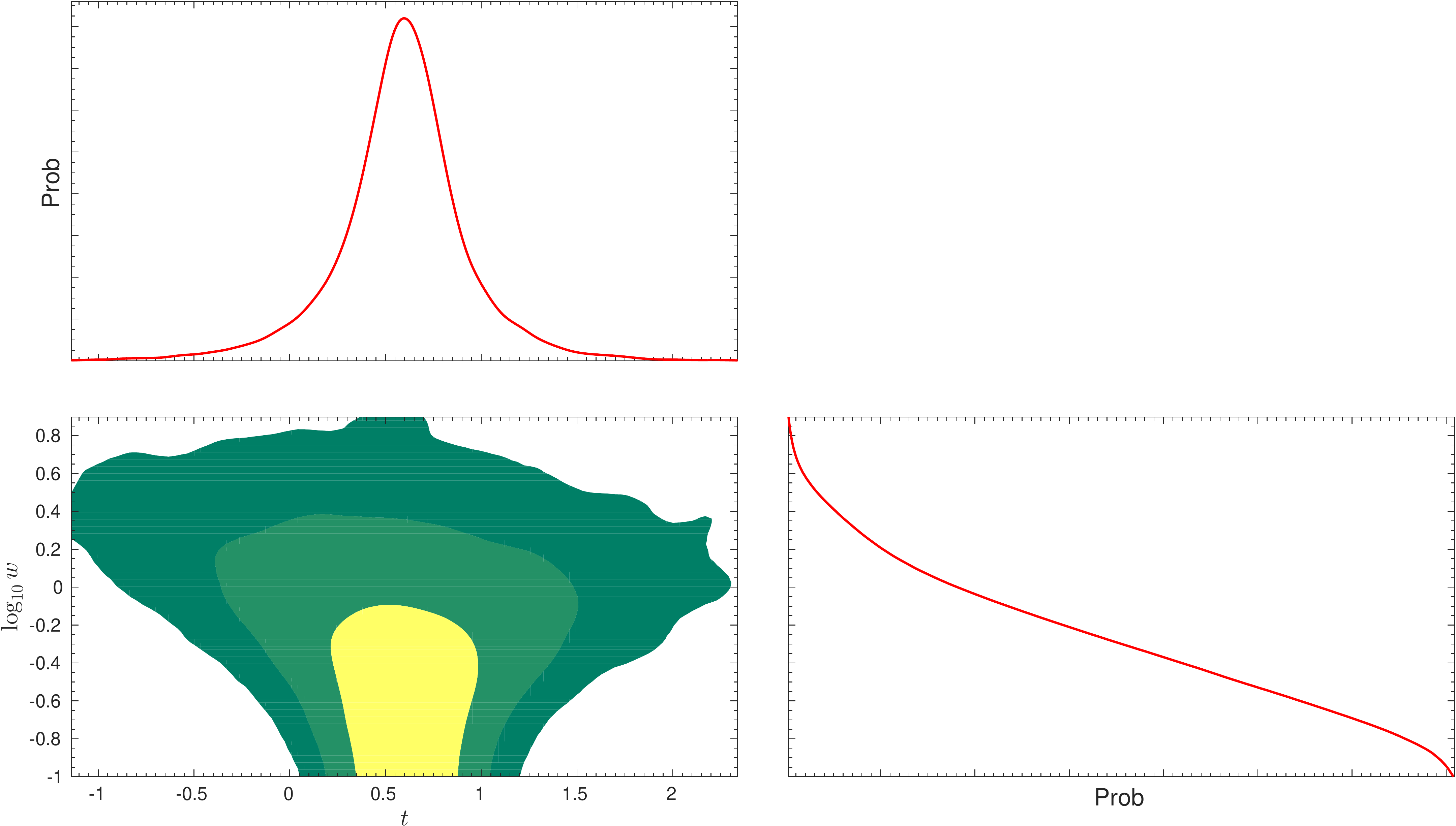}
  \caption{Results of the solar neutrino analysis for the generalized
    SSM.  The lower left panel shows the two-dimensional
    iso-likelihood contours at $1\sigma$, $2\sigma$ and $3\sigma$ in
    the plane $(t,\omega)$.  The upper left and lower right panels
    show the one-dimensional probability distributions for $t$ and
    $\omega$ respectively.}
  \label{fig:SSMext_tri}
\end{figure}

We plot the results of this generalized-SSM analysis in
Fig.~\ref{fig:SSMext_tri} where we show the two-dimensional
iso-likelihood contours for $1\sigma$, $2\sigma$ and $3\sigma$ in the
plane $(t,\omega)$ as well as the one-dimensional probability
distributions for each parameter.  From the upper panel we see that a
model with $t\simeq 0.6$ is presently favored by the data, and
provides a description which is clearly better than the limiting cases
of the AGSS09 and GS98 models at $t=0$ and $t=1$ (characterized by
rather similar probability as expected from the previous
section). Also looking at the bi-dimensional region we see that this
is more the case when allowing for smaller theoretical uncertainties
than presently given in the SSM predictions, \textit{i.e}, the minimum
likelihood lies at values of $\omega<1$. The two-dimensional regions
present a ``funnel'' shape at lower $\omega$ because
$\sigma_i^\textsc{ssm}$ becomes much smaller than
$\sigma_i^\textsc{fit}$ and therefore the analysis becomes independent
of $\omega$. The fact that a better description of the neutrino data
is obtained for a model with reduced theoretical uncertainties
indicates that even with the present neutrino data some refinement on
the models can be obtained by including the results of the solar
neutrino data as inputs in the model
construction~\cite{inpreparation}.

\section{Summary and outlook}
\label{sec:sum}

The pioneering proposal of using neutrinos to verify the source of the
energy produced in the Sun has ended in the discovery of flavor
conversion among solar neutrinos and in quantifying the contribution
of the main mechanism of energy generation in the Sun. Further
progress is needed to precisely answer some fundamental questions in
solar evolution, such as (i) how much constrained are non-standard
sources of energy, (ii) how much the CNO mechanism contributes to the
solar energy generation, and (iii) what is the solution to the solar
abundances problem.

In this work, we have updated the determination of solar model
independent neutrino fluxes presented in
Ref.~\cite{GonzalezGarcia:2009ya} by taking into account the latest
data from both solar and non-solar neutrino experiments. We have
derived the best neutrino oscillation parameters and solar fluxes
constraints using a Bayesian analysis with and without imposing
nuclear physics as the only source of energy generation (luminosity
constraint).

The precise measurement of the rate of \Nuc[7]{Be} solar neutrinos by
the Borexino experiment~\cite{Bellini:2011rx, Bellini:2013lnn}
together with their first direct detection of \Nuc{pp}
neutrinos~\cite{Bellini:2014uqa} and the very precise measurement of
the mixing angle $\theta_{13}$ greatly contribute to answer the first
question and constrain non-standard sources of energy, other than
nuclear physics, as shown in Eq.~\eqref{eq:lnutot}. The uncertainty on
the total luminosity due to nuclear physics derived from neutrino data
has been reduced by a factor two and is now, for the first time, below
10\%.

Present data cannot yet answer the second and third questions. The
discovery of CNO neutrinos is within reach of the existing liquid
scintillator detectors, if sufficient level of purification could be
achieved. We have shown that present bounds on CNO neutrino fluxes are
very close to the theoretical $3\sigma$ range, whether or not other
sources of energy contribute to the energy generation. A discovery
would not only verify the main mechanism of energy generation for
bigger (or older) stars than our Sun, it would also help to solve the
solar abundances problem.  We have shown that a CNO flux measurement
with $\sigma_\textsc{cno} = 5\%$ uncertainty can lead to a moderate
evidence in favor of one of the two alternative sets of solar
abundances. Either the abundances are larger than what the most
refined determinations indicate, or the opacities and stellar
evolution codes have to be revisited to fit the precise
helioseismology observations.

\section*{Acknowledgments}

This work is supported by Spanish MINECO grants ESP2014-56003-R,
FPA2011-29678, FPA2012-31880, FPA2012-34694 and FPA2013-46570, by the
Severo Ochoa program SEV-2012-0249 of IFT, by the Maria de Maeztu
program MDM-2014-0369 of ICCUB, and consolider-ingenio 2010 grant
CSD-2008-0037, by CUR Generalitat de Catalunya grants 2014-SGR-104 and
2014-SGR-1458, by Generalitat Valenciana Prometeo grant II/2014/050,
by USA-NSF grant PHY-13-16617, and by EU grant FP7 ITN INVISIBLES
(Marie Curie Actions PITN-GA-2011-289442).

\appendix

\section{Borexino}
\label{sec:app-borex}

Our analysis of the \Nuc{pp} neutrino signal recently observed by
Borexino is entirely based on the information provided
in~\cite{Bellini:2014uqa}. The set of operations which we have
performed in order to gain confidence with such data can be broadly
divided into two parts.  First of all, we have focused solely on
reproducing their fit, which involves extracting the information from
the paper and ensuring that we can handle it properly. In this part we
define:
\begin{equation} \label{eq:borex}
  N_b^\text{th}(\vec\xi) =
  N_b^\text{sun}(\vec\xi) + N_b^\text{bkg}(\vec\xi)
  \quad\text{with}\quad
  \left\lbrace
  \begin{aligned}
    N_b^\text{sun}(\vec\xi)
    &= \sum_f N_{b,f}^\text{sun}
    \big( 1 + \pi_f^\text{sun} \xi_f^\text{sun} \big) \,,
    \\
    N_b^\text{bkg}(\vec\xi)
    &= \sum_i N_{b,i}^\text{bkg}
    \big( 1 + \pi_i^\text{bkg} \xi_i^\text{bkg} \big)
  \end{aligned}
  \right.
\end{equation}
where $\vec\xi$ is a set of variables parametrizing the theoretical
and systematic uncertainties. Here $b \in \lbrace 1,\, \dots,\, 158
\rbrace$ identifies the data bin, $f \in \lbrace \Nuc{pp},\,
\Nuc[7]{Be},\, \Nuc{pep},\, \text{CNO} \rbrace$ is the solar flux, and
$i \in \lbrace \Nuc[14]{C},\, \Nuc[85]{Kr},\, \Nuc[210]{Bi},\,
\Nuc[210]{Po},\, \Nuc[214]{Pb},\, \text{pile-up} \rbrace$ labels the
background component. Following Refs.~\cite{Bellini:2014uqa,
  Smirnov:2015lxy} we define the priors $\pi_f^\text{sun}$ and
$\pi_i^\text{bkg}$ as follows:
\begin{equation}
  \begin{aligned}
    \text{fixed:} & \quad
    \pi_{\Nuc{pep}}^\text{sun} = \pi_\textsc{cno}^\text{sun}
    = \pi_{\Nuc[214]{Pb}}^\text{bkg} = 0 \,,
    \\
    \text{constrained:} & \quad
    \pi_{\Nuc[7]{Be}}^\text{sun} = 2.3/48 \,,
    \quad
    \pi_{\Nuc[14]{C}}^\text{bkg} = 1/40 \,,
    \quad
    \pi_\text{pile-up}^\text{bkg} = 7/321 \,,
    \\
    \text{free:} & \quad
    \pi_{\Nuc{pp}}^\text{sun} = \pi_{\Nuc[85]{Kr}}^\text{bkg}
    = \pi_{\Nuc[210]{Po}}^\text{bkg} = \pi_{\Nuc[210]{Bi}}^\text{bkg}
    \to \infty \,.
  \end{aligned}
\end{equation}
We have extracted both the solar neutrino fluxes and the backgrounds
from the upper panel of Fig.~3 of Ref.~\cite{Bellini:2014uqa}. We have
converted these spectra into absolute number of events for each bin
$b$ (for the solar flux and the background ) by multiplying the given
event rates (c.p.d.\ per 100~t per keV) by the total data-taking time
($T^\text{run} = 408$~days), the fiducial volume (75.47~t), and the
specific bin energy size.  We have verified that the sum of the
different contributions agrees reasonably well (within the resolution
of the figure) with the ``best-fit prediction'' shown as a black solid
line in the figure. We have taken care to rescale the \Nuc[14]{C} and
the \Nuc[7]{Be} spectra extracted from Ref.~\cite{Bellini:2014uqa} by
$40/39.8$ and $48/46.2$, respectively, to match the priors quoted in
Sec.~3.4 of Ref.~\cite{Smirnov:2015lxy}.

In order to test our ability to reproduce the Borexino fit, we have
constructed a $\chi^2$ function as follows:
\begin{equation}
  \chi^2 = \min_{\vec\xi} \left\lbrace \sum_b
  \frac{\big[ N_b^\text{th}(\vec\xi) - N_b^\text{ex} \big]^2}{N_b^\text{ex}}
  + \sum_f \big( \xi_f^\text{sun} \big)^2
  + \sum_i \big( \xi_i^\text{bkg} \big)^2
  \right\rbrace \,.
\end{equation}
Here $N_b^\text{ex}$ is the observed number of events for the bin $b$,
which we have derived from the residuals $\rho_b$ shown in the lower
panel of Fig.~3. Note that, lacking the information on possible
correlations among different bins, we have assumed that the
experimental data are uncorrelated and that the statistical error is
simply the square root of the number of events, which implies
$\sqrt{N_b^\text{ex}\vphantom{N^\text{th}}} = \rho_b/2 +
\sqrt{(\rho_b/2)^2 + N_b^\text{th}}$. We have then performed a fit of
the various spectra against the experimental data, and we have
verified that the best-fit values and allowed ranges which we obtain
(both solar fluxes and backgrounds) are in excellent agreement with
those listed above Fig.~3. This proves that our simplified approach is
credible and ensures a realistic determination of the solar flux
normalizations, which is the main topic of this work.

\bigskip

The second step of our procedure requires embedding this fit into our
global analysis in a consistent way, and making sure that its accuracy
is not spoiled. To this aim, we now discard the solar spectra
$N_b^\text{sun}(\vec\xi)$ previously introduced in
Eq.~\eqref{eq:borex} and define instead:
\begin{equation} \label{eq:ourown}
  N_b^\text{th}(\vec\omega, \vec\xi) =
  n_\text{el} T^\text{run} \sum_\alpha \int
  \frac{d\Phi_\alpha^\text{det}}{dE_\nu}(E_\nu | \vec\omega) \,
  \frac{d\sigma_\alpha}{d T_e}(E_\nu, T_e) \,
  R_b(T_e | \vec\xi) \, dE_\nu +
  N_b^\text{bkg}(\vec\xi) \,.
\end{equation}
Note that the backgrounds $N_b^\text{bkg}(\vec\xi)$ are the same as
before. In Eq.~\eqref{eq:ourown} $\vec\omega$ describes both the
neutrino oscillation parameters and the eight solar flux
normalizations, $n_\text{el}$ is the number of electron targets,
$d\sigma_\alpha / dT_e$ is the elastic scattering differential
cross-section for neutrinos of type $\alpha \in \lbrace e, \mu, \tau
\rbrace$, and $d\Phi_\alpha^\text{det} / {dE_\nu}$ is the
corresponding flux of solar neutrinos \emph{at the detector} --~hence
it incorporates the neutrino oscillation probabilities. For comparison
with the Borexino results we have used a three-neutrino oscillation
model with values $\sin^2\theta_{13} = 0.022$, $\sin^2\theta_{13} =
0.304$ and $\Dmq_{21} = 7.5\times 10^{-5}~\eVq$ for the relevant
parameters, and assumed the GS98 solar model.

The detector response function $R_b(T_e | \vec\xi)$ depends on the
\emph{true} electron kinetic energy $T_e$ as well as three new
systematic variables $\xi_\text{vol}$, $\xi_\text{scl}$ and
$\xi_\text{res}$ which we have included for completeness and
consistency with the simulations of other experiments:
\begin{equation}
  R_b(T_e | \vec\xi) = (1 + \pi_\text{vol}\, \xi_\text{vol})
  \int_{T_b^\text{min}(1 + \pi^b_\text{scl}\, \xi_\text{scl})}
  ^{T_b^\text{max}(1 + \pi^b_\text{scl}\, \xi_\text{scl})}
  \Gauss\left[ T_e - T',\, \sigma_T
    (1 + \pi_\text{res}\, \xi_\text{res}) \right] \, dT' \,.
\end{equation}
Here $\Gauss(x, \sigma) \equiv \exp\left[- x^2 / 2\sigma^2 \right] /
\sqrt{2\pi} \sigma$ is the normal distribution function, while
$T_b^\text{min}$ and $T_b^\text{max}$ are the boundaries of the
\emph{reconstructed} electron kinetic energy $T'$ in the bin $b$. We
have assumed an energy resolution $\sigma_T / T_e = 5.5\% /
\sqrt{T_e\, \text{[MeV]}}$, a fiducial volume uncertainty
$\pi_\text{vol} = 2\%$, an energy scale uncertainty $\pi_\text{scl} =
1\%$, and an arbitrary energy resolution uncertainty $\pi_\text{res} =
5\%$, all uncorrelated between Borexino Phase I and Phase II.

As a first check, we have explicitly verified that our first-principle
calculation of the solar flux contribution to the various bins matches
quite accurately the $N_{b,f}^\text{sun}$ spectra extracted from
Fig.~3 of Ref.~\cite{Bellini:2014uqa}. We have then constructed a new
$\chi^2$ function for Borexino Phase II:
\begin{equation}
  \chi^2(\vec\omega) = \min_{\vec\xi} \left\lbrace \sum_b
  \frac{\big[ N_b^\text{th}(\vec\omega, \vec\xi) - N_b^\text{ex} \big]^2}{N_b^\text{ex}}
  + \sum_i \big( \xi_i^\text{bkg} \big)^2
  + \xi_\text{vol}^2
  + \xi_\text{scl}^2
  + \xi_\text{res}^2
  \right\rbrace
\end{equation}
and we have verified once more that our final fit (after combining it
with the Borexino Phase I data to provide a prior for the \Nuc[7]{Be}
flux) still yields the correct best-fit values and allowed ranges for
both the \Nuc{pp} solar flux normalization and the Borexino
backgrounds. Thus we consider that our proposed goal, namely to embed
Borexino \Nuc{pp} data into our codes in a realistic and consistent
way, has been accomplished.

\begin{figure}\centering
  \includegraphics[width=\textwidth]{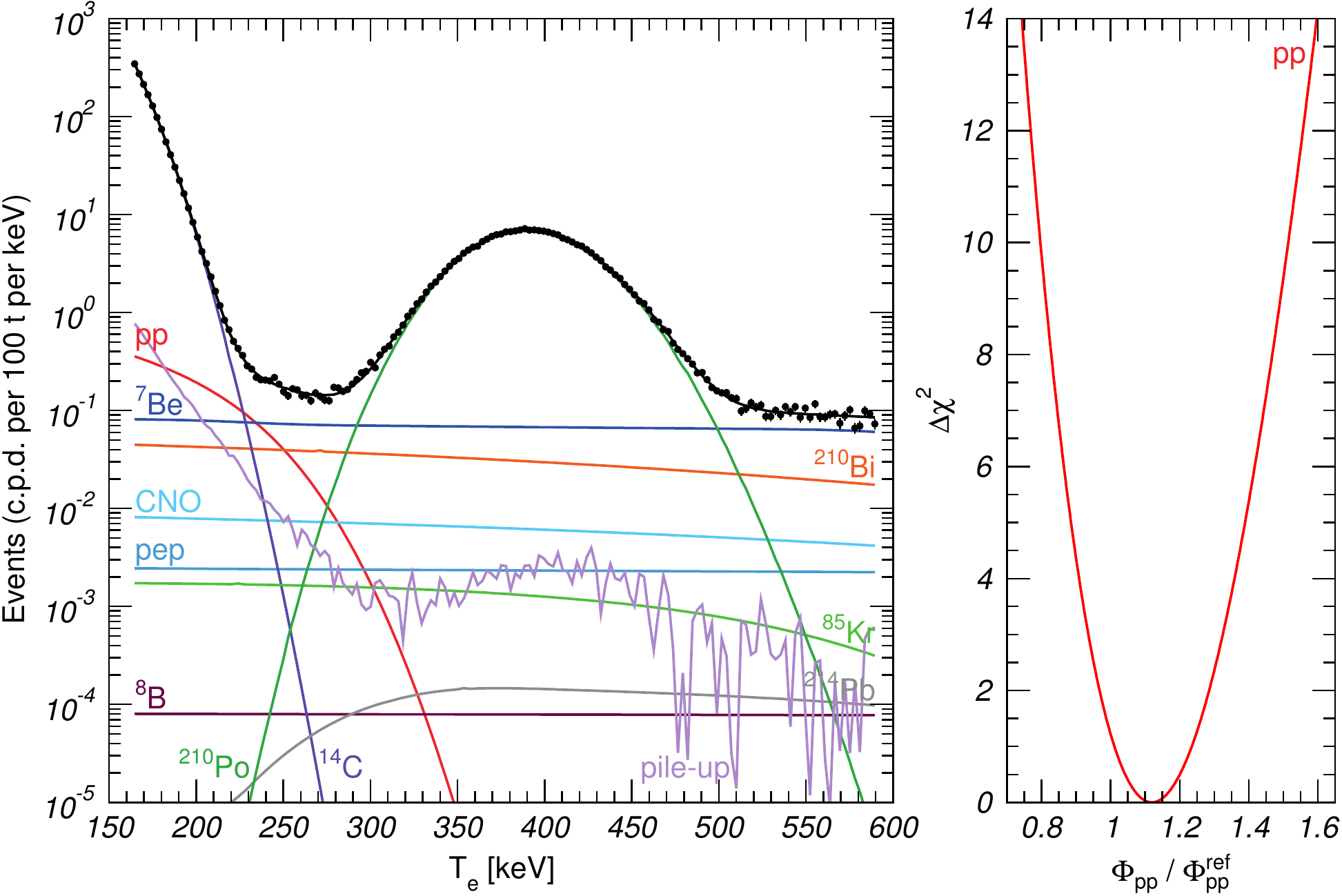}
  \caption{Spectrum for the best fit point of our spectral fit to the
    Borexino Phase II data in the energy region between 165--590 keV
    under the assumptions described in the Appendix (left), and
    $\Delta\chi^2$ as a function of the \Nuc{pp} flux (right).}
  \label{fig:borexino}
\end{figure}

In Fig.~\ref{fig:borexino} we show the results of our
analysis. Comparing the left panel with Fig.~3 of
Ref.~\cite{Bellini:2014uqa} we observe a very good agreement in the
best fit determination of both solar fluxes and backgrounds, as
mentioned above.  In particular, the allowed range for
$\Phi_{\Nuc{pp}}$ is perfectly compatible with the value
$\Phi_{\Nuc{pp}} = (6.6 \pm 0.7) \times 10^{10} \,\text{cm}^{-2}
\,\text{s}^{-1}$ quoted by the Borexino collaboration, as can be seen
from the right panel where we plot the $\Delta\chi^2$.

\bibliographystyle{JHEP}
\bibliography{references}

\end{document}